\documentclass[lettersize,journal]{IEEEtran}
\usepackage{amsmath,amsfonts}
\usepackage{algorithmic}
\usepackage{algorithm}
\usepackage{array}
\usepackage[caption=false,font=normalsize,labelfont=sf,textfont=sf]{subfig}
\usepackage{textcomp}
\usepackage{soul}
\usepackage{xcolor}
\usepackage{tcolorbox}
\usepackage[normalem]{ulem}
\usepackage{adjustbox}
\usepackage{mathtools}
\usepackage{stfloats}
\usepackage{url}
\usepackage{romannum}
\usepackage{verbatim}
\usepackage{graphicx}
\usepackage{cite}
\usepackage{tabularx}
\usepackage{makecell}
\usepackage{multirow}
\definecolor{mycolor}{rgb}{1,1,0}

\begin{document}

\title{Realistic Channel and Delay Coefficient Generation for Dual Mobile Space-Ground Links -- A Tutorial}

\author{Hongzhao Zheng,~\IEEEmembership{Member,~IEEE}, Mohamed Atia,~\IEEEmembership{Senior Member,~IEEE}, Halim Yanikomeroglu,~\IEEEmembership{Fellow,~IEEE}
\thanks{H. Zheng is with the Embedded and Multi-sensor Systems Lab (EMSLab) and the Non-Terrestrial Networks (NTN) Lab, Department of Systems and Computer Engineering, Carleton University, Ottawa, ON, K1S 5B6, Canada (e-mail: hongzhaozheng@cmail.carleton.ca).}
\thanks{M. Atia is with the Embedded and Multi-sensor Systems Lab (EMSLab), Department of Systems and Computer Engineering, Carleton University, Ottawa, ON, K1S 5B6, Canada (e-mail: Mohamed.atia@carleton.ca).}
\thanks{H. Yanikomeroglu is with the Non-Terrestrial Networks (NTN) Lab, Department of Systems and Computer Engineering, Carleton University, Ottawa, ON, K1S 5B6, Canada (e-mail: halim@sce.carleton.ca).}
}

\markboth{}%
{Shell \MakeLowercase{\textit{et al.}}: A Sample Article Using IEEEtran.cls for IEEE Journals}



\maketitle

\begin{abstract}
Channel and delay coefficient are two essential parameters for the characterization of a multipath propagation environment. It is crucial to generate realistic channel and delay coefficient in order to study the channel characteristics that involves signals propagating through environments with severe multipath effects. While many deterministic channel models, such as ray-tracing (RT), face challenges like high computational complexity, data requirements for geometrical information, and inapplicability for space-ground links, and nongeometry-based stochastic channel models (NGSCMs) might lack spatial consistency and offer lower accuracy, we present a scalable tutorial for the channel modeling of dual mobile space-ground links in urban areas, utilizing the Quasi Deterministic Radio Channel Generator (QuaDRiGa), which adopts a geometry-based stochastic channel model (GSCM), in conjunction with an International Telecommunication Union (ITU) provided state duration model. This tutorial allows for the generation of realistic channel and delay coefficients in a multipath environment for dual mobile space-ground links. We validate the accuracy of the work by analyzing the generated channel and delay coefficient from several aspects, such as received signal power and amplitude, multipath delay distribution, delay spread and Doppler spectrum.  

\end{abstract}


\section{Introduction}
\IEEEPARstart{V}{ertical} heterogeneous networks (VHetNets) have garnered significant interest in the community due to its potential to offer enhanced coverage and capacity, improved spectrum utilization, lower latency, and support for diverse use cases \cite{b34}. As such it is seen as a crucial facilitator in the forthcoming era of 6G and beyond for a variety of usage scenarios, such as extremely reliable and low latency communications (ERLLC), ultra-massive machine-type communications (umMTC), long-distance and high-mobility communications (LDHMC), and further enhanced mobile broadband (FeMBB) \cite{b31,b32,b33}. As researchers explore the potential advantages and practical use cases of integrating diverse vertical heterogeneous infrastructures, it is essential to approach many research findings with caution due to the absence of realistic signals in the scenarios they are investigating. These scenarios encompass signals such as millimeter-wave (mmWave), terahertz (THz), and high altitude platform station (HAPS) signals, spanning across space, air, ground, deep sea, rural, and urban environments. This lack of realistic signals necessitates numerous studies to focus primarily on the link level rather than encompassing the broader system level. 

With the increasing emergence of standardization efforts, technical specifications, and reports from organizations such as the International Telecommunication Union (ITU) and the 3rd Generation Partnership Project (3GPP), which include channel models and line-of-sight (LOS) probability models, there is growing interest within the community to develop a comprehensive platform that enables the customization of various parameters.  These include antenna attributes like type, transmit/receive gain, pattern, and dimensions, as well as vertical object placements, among others. Such a platform would allow for an efficient and accurate generation of realistic channel coefficients, time delays, and even pseudo signals for the target infrastructures.

In general, there are two types of channel models in wireless communications: deterministic and stochastic. Each of them comes with its own set of strengths and weaknesses. For instance, as deterministic channel models are concerned with an accurate modeling of the wave propagation by solving Maxwell's equations, they are expected to provide higher levels of accuracy. However, deterministic channel models are specific to locations and demand detailed information about the geometry and electromagnetic (EM) characteristics of the propagation environment, leading to increased computational complexity \cite{b27,b28}. One commonly employed deterministic channel model is the ray-tracing (RT) model \cite{b2,b3,b4}. However, this model often requires a huge amount of data including but not limited to a high definition map of the environment and surface parameters of the buildings, which brings not only the difficulty in data access but also high computational cost \cite{b23,b24}. Additionally, the RT method is generally not applicable for space-ground links due to the long distance involved \cite{b25}. 

On the other hand, stochastic channel models primarily depend on the statistics of the key parameters of the wireless channel, such as the number of multipath components (MPCs), azimuth and elevation spread  of arrivals and departures, Ricean K-factor, shadow fading and so forth. Without the need for the detailed information about the environment, stochastic channel models tend to exhibit reduced computational complexity but offer relatively lower accuracy compared to deterministic channel models \cite{b36}. Broadly, stochastic channel models can be classified into two types: geometry-based stochastic channel models (GSCMs) and nongeometry-based stochastic channel models (NGSCMs). Similar to deterministic channel models, GSCMs consider the geometrical component, such as scatterers. However, GSCMs differ in that the locations of the scatterers are selected stochastically rather than deterministically based on an environment database \cite{b29}. In contrast to GSCMs, NGSCMs entirely rely on statistical parameters in modeling the wireless channel. As a result, NGSCMs are unable to provide spatial consistency \cite{b27} and accurate physical insights of the system, leading to lower accuracy when applied to different scenarios unless the site-specific statistics are available \cite{b30}.

To achieve a balance between computational complexity and accuracy, the Quasi Deterministic Radio Channel Generator (QuaDRiGa), which follows the GSCM approach, is one of the stochastic models that has been widely adopted for channel generation. This model is originally developed based on the Wireless World Initiative for New Radio (WINNER) channel model proposed in WINNER \Romannum{2} deliverable D1.1.2 v1.1 \cite{b5} and is now known as a 3GPP-3D and 3GPP 38.901 reference implementation \cite{b6}. It offers features such as continuous time evolution, spatially correlated large and small-scale fading, transitions between varying propagation scenarios, and so on. QuaDRiGa is also considered a statistical ray-tracing model, as such it can be used to effectively and efficiently model multipath, channel fading, and Doppler effect. Due to the mechanism of the QuaDRiGa channel generator, this tutorial is expected to be applicable for the dual mobile air-ground links, which facilitates a realistic simulation for an integrated signal processing, such as urban positioning \cite{b7}, mmWave communications, THz communications \cite{b8}, and multiple-input multiple-output (MIMO) communications \cite{b9} in a vertical heterogeneous network (VHetNet)\cite{b10}. We clarify that QuaDRiGa currently lacks support for THz communications due to inadequate measurements. Nonetheless, it is anticipated that QuaDRiGa will expand into the THz band once sufficient measurements become available. 

Although QuaDRiGa provides numerous tutorials for different channel simulations, none of them takes into account the realistic state transitions that occur in the real world. Typically, a state, either LOS or non-line-of-sight (NLOS), would persist for a certain duration before transitioning to the other state. To accurately model the signal propagation in a multipath environment, it is therefore necessary to consider a realistic state transition, especially when a long period of simulation is desired or synchronization problems are of interest. A traditional approach to model the state transition is by employing a Markov model, wherein the subsequent state is determined by a transition probability model without considering the past occurrences. However, due to the fact that the channel properties are assumed to be quasi-stationary only for a short period of time and that state durations follow an exponential distribution, the Markov models may yield unrealistic state durations \cite{b15}. Therefore, the statistical state duration model that follows a semi-Markov approach has gained a greater acceptance in embodying the practical duration of different states. This approach has received endorsement from the ITU and is documented in ITU-R P.681-11 \cite{b13}.

Space-ground links involve the communication between satellites in space and ground-based receivers, while air-ground links pertain to the communication between aircraft or drones and ground-based receivers. Both links cover substantial distances and share similar propagation characteristics and challenges. The insights gained from studying space-ground links, including signal propagation, attenuation, multipath effects, and interference, can be applied effectively to model air-ground links. Moreover, the methodologies used for space-ground link modeling can be adapted and extended for air-ground link analysis. On the other hand, space-ground links present additional challenges, such as longer propagation distances, orbital dynamics, and significant Doppler effects, particularly in cases where both terminals are in motion. Consequently, it is more rational to begin by modeling the scenario involving dual mobile space-ground links and subsequently extend these models to air-ground links.

Multipath propagation has consistently posed a noteworthy obstacle in wireless communications, particularly in demanding conditions like indoor spaces, wooded regions, and urban environments. In these settings, signals become more vulnerable to phenomena like reflections, refractions, and scattering. Due to multipath, multiple copies of the same signal arrive at the receiver with different time delays and amplitudes, posing many issues such as intersymbol interference (ISI), signal fading, and degraded signal quality\cite{b1}. Hence, it is crucial to develop a platform that can emulate a system which incorporates signal processing in a multipath environment like urban areas. 

In this paper, we have developed a scalable tutorial that enables the generation of realistic channel and delay coefficients for dual mobile space-ground links in urban areas. We utilize the combined capabilities of the Matlab Satellite Communications Toolbox, Skydel GNSS Simulator, and QuaDRiGa channel generator. While QuaDRiGa provides tutorials for dual mobile ground-ground links and single mobile space-ground links, simulating scenarios involving dual mobile space-ground links presents additional challenges. For example, a cautious manipulation of the spatial coordinates of satellites with respect to the receiver is required in consideration of the Earth curvature and coordinate transformation. Moreover, the choice of channel update rate should be carefully considered in order to accurately capture the Doppler effect caused by the rapid movement of satellites. To examine the accuracy of the tutorial, we analyze the generated channel and delay coefficients from a variety of angles, such as the received signal power and amplitude, multipath delay distribution, delay spread, and Doppler spectrum. Due to the unavailability of actual GPS data that aligns with our statistical model, we compare certain aspects of our work with the findings of others who have access to real GPS measurements. This work serves as a tutorial for potential QuaDRiGa users and enhances the available resources for simulating dual mobile space-ground links with time-varying channels. The key contributions of this paper are listed as follows:

\begin{enumerate}
    \item In the preparation of terminal trajectories with designated propagation scenarios, we integrate a practical phenomenon in signal reception by employing both a LOS probability model and a state transition model and following the semi-Markov approach.

    \item By leveraging the combined capabilities of the QuaDRiGa channel generator, the Matlab Satellite Communications Toolbox, and the Skydel GNSS simulator, a tutorial for dual mobile space-ground links has been meticulously crafted.

    \item To serve the purpose as a tutorial, we outline several challenges that we encounter in the channel modeling for the scenario we consider. In addition, we point out a few possible mistakes/typos in the source code provided by QuaDRiGa.
    
    \item To verify the accuracy of the generated channel and delay coefficient, we provide a comprehensive analysis on the received signal power and amplitude, multipath delay distribution, delay spread, and Doppler spectrum. The results of this analysis show a high degree of agreement with previous researches that incorporate real GPS measurements.

\end{enumerate}

The rest of the paper is organized as follows: The simulation descriptions, such as satellite scenario, receiver position, and channel modeling are introduced in Section \Romannum{2}. The challenges encountered in developing this tutorial as well as the possible mistakes and typos in the source code provided by QuaDRiGa are enumerated in Section \Romannum{3}. The analytical metrics used to examine the fidelity of the channel and delay coefficient are described in Section \Romannum{4}. A brief description of the simulation setup is given in Section \Romannum{5}. Section \Romannum{6} presents the detailed analysis of the simulation results concerning the received signal power and amplitude, multipath delay distribution, delay spread, and Doppler spectrum. Finally, the conclusion, the implications of the work, and an outline of the future work are provided in Section \Romannum{7}. 

This manuscript serves an instructional purpose, and as such, it is structured to facilitate focused study in specific areas of interest. Those seeking insights into received signal power and amplitude are directed to Section IV-A and VI-A. For an understanding of multipath delay distribution, Section IV-B and VI-B should be referenced. Inquiries regarding delay spread are comprehensively addressed in Section IV-C and VI-C. Finally, for readers with a focus on the Doppler spectrum, Section IV-D and VI-D offer detailed exposition and analysis.

\section{Simulation Descriptions}
In this section, the descriptions on the channel modeling using the QuaDRiGa channel generator, the creation of satellite scenarios using the Matlab Satellite Communications Toolbox, and the generation of receiver positions using the Skydel GNSS simulator are presented.

\subsection{Channel Modeling - QuaDRiGa}\label{AA}
In order to utilize QuaDRiGa channel generator effectively, users are expected to provide three essential inputs. These inputs encompass terminal trajectories assigned with specific propagation scenarios, a network layout that outlines transmitter positions and includes the carrier frequency information, as well as the necessary antenna parameters. A terminal trajectory can be divided into multiple segments, each of which needs to be assigned with a propagation environment. There are many propagation scenarios prepared as configuration files provided by QuaDRiGa, such as the non-terrestrial-network (NTN) urban LOS, 3GPP 38.901 urban macrocell (UMa) NLOS, 3GPP 37.885 highway LOS, and so on. The scenarios supported by QuaDRiGa include 3GPP NR UMa, 3GPP NR urban microcell (UMi), highway, indoor, rural macrocell (RMa), rural microcell (RMi), MIMO communications, NTN communications, and so forth. In this work, we employ the NTN urban LOS and NTN urban NLOS. With these two propagation scenarios, we can simulate scenarios not only for satellites but also for other vertical infrastructures such as HAPS and unmanned aerial vehicles (UAVs). 

\begin{figure}[t]
\centerline{\resizebox{\columnwidth}{!}{\includegraphics{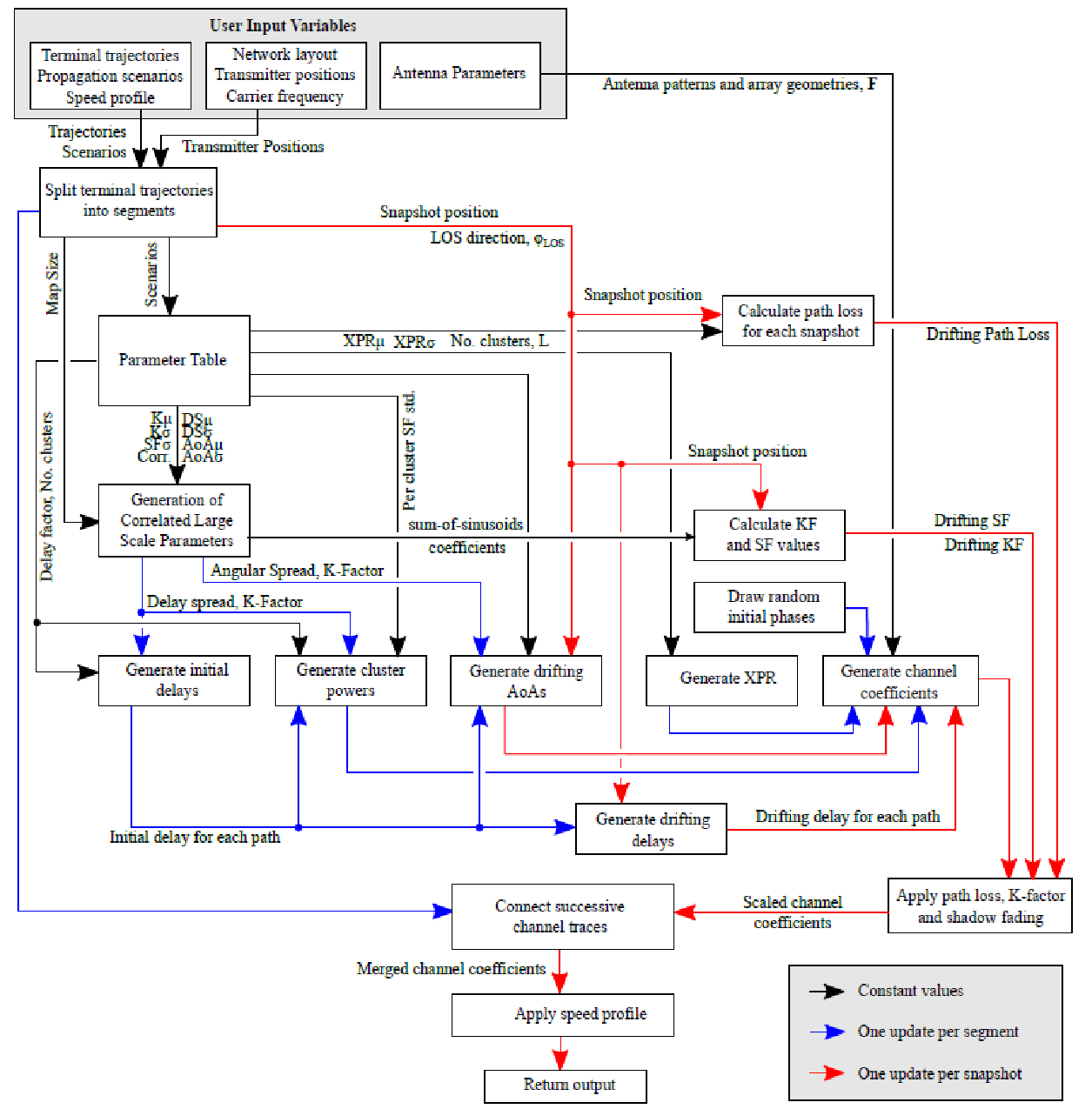}}}
\caption{Data flow of QuaDRiGa channel model (parameters on black lines are constant, they are either provided by the user or the model; parameters on blue lines are updated once per segment; parameters on red lines are updated once per snapshot) \cite{b6}.}
\label{fig2}
\end{figure}

It is important to note that in QuaDRiGa, the LOS scenario incorporates both the direct path and MPCs, whereas in the NLOS scenario, only MPCs are considered. For each segment the channel parameters, such as delay spread, angle spread, shadow fading, cross-polarization ratio and so on, are stochastically determined based on the statistical distributions derived from channel measurement campaigns \cite{b5,b14}. In addition, a number of scattering clusters are generated based on the corresponding channel parameters and the receiver position in each segment. There are 20 sub-paths associated with each cluster. The arrival angles are calculated for each sub-path and for each position of the receiver along the trajectory. Based on the antenna parameters assigned to the transmitter (TX) and the receiver (RX), the complex-valued coefficients that contain phase information for all sub-paths are generated. Finally, the channel coefficients and the path delays are generated by summing up the coefficients of sub-paths and merging the adjacent segments. This approach facilitates the observation of realistic phenomena, such as constructive and destructive interference, within the generated channel and delay coefficients. The data flow of QuaDRiGa channel model is shown in Fig.~\ref{fig2}.

To emulate a realistic LOS/NLOS conditions for satellites, the LOS probability model proposed in \cite{b11} and \cite{b12}, which is a function of satellite elevation angles, is implemented to determine the visibility condition for each satellite at the very first epoch. The QuaDRiGa channel generator automatically simulates multipath and channel fading. Based on this mechanism, we categorize the propagation state into two categories: GOOD and BAD. A GOOD state includes both a direct path and MPCs, while a BAD state only involves the MPCs. The state transition is implemented by following the semi-Markov approach proposed in \cite{b13}. There is a potential error in this document, wherein a factor of 2 is omitted under the square root in the denominator, and there is a minus sign missing in the exponent. As a result, the equation is corrected as follows:
\begin{equation}
p_{\text{lognormal}}(x) = \frac{1}{\sigma_i x \sqrt{2\pi}} \exp(-\frac{(\ln(x)-\mu_i)^2}{2\sigma_i^2}) \label{eq1}
\end{equation}

\noindent where $i$ denotes the binary state which can either be GOOD or BAD, $\mu_i$ denotes the mean duration in meter for state $i$, $\sigma_i$ denotes the standard deviation of the duration in meter for state $i$. Both $\mu_i$ and $\sigma_i$ are scenario specific and are provided in Annex 2 of \cite{b13}. In order to simulate a realistic state duration, it is essential to consider the scenario-specific minimum state duration, denoted as $dur_{(\mathrm{min})_i}$, which is dependent on the elevation angle $\alpha$ as mentioned in [13]. The summarized values of the minimum state duration are presented in Table \Romannum{1}. 

\begin{table}[t]
\caption{Minimum State Duration}
\begin{center}
\begin{tabularx}\linewidth{>{\centering\arraybackslash}X|>{\centering\arraybackslash}X|>{\centering\arraybackslash}X}
\Xhline{1pt}
\multicolumn{3}{c}{\textbf{\centering Frequencies between 1.5 and 3 GHz}}\\
\Xhline{1pt}
$\alpha$ [degrees]& $\text{dur}_{(\text{min})_{\text{GOOD}}}$ [m]& $\text{dur}_{(\text{min})_{\text{BAD}}}$ [m]\\
\hline
20 & 3.9889 & 10.3114\\
\hline
30 & 7.3174 & 5.7276\\
\hline
45 & 10.0 & 6.0\\
\hline
60 & 10.0 & 1.9126\\
\hline
70 & 118.3312 & 4.8569\\
\Xhline{1pt}
\end{tabularx}
\label{tab1}
\end{center}
\end{table}

\subsection{Satellite Scenario - Matlab Satellite Communications Toolbox}
\begin{figure}[t]
\centerline{\resizebox{\columnwidth}{!}{\includegraphics{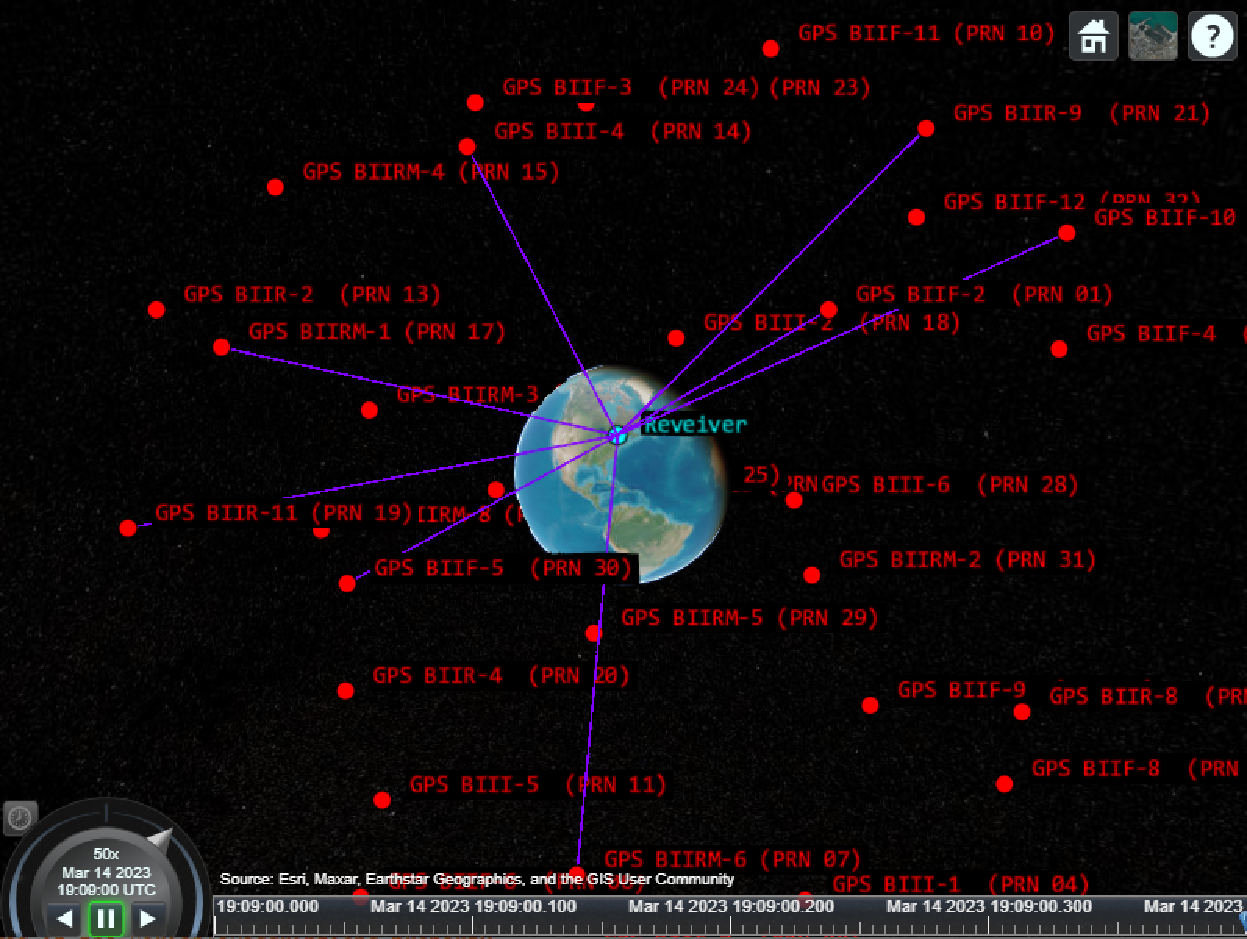}}}
\caption{A snapshot of the satellite scenario simulated using the Matlab Satellite Communications Toolbox.}
\label{fig1}
\end{figure}

The position of the transmitter or satellite constitutes a crucial input for the QuaDRiGa channel generator. Ensuring utmost precision in the system dynamics necessitates an accurate simulation of satellite positions across various time epochs. To facilitate this, the Matlab Satellite Communications Toolbox offers tools that adhere to established standards. These tools enable an efficient creation of a comprehensive satellite scenario by specifying simulation start and end times, pinpointing the receiver or ground station location, and supplying a two-line element (TLE) file that encompasses a compilation of the orbital elements for operational GPS satellites. From this satellite scenario, we can gather information such as satellite positions over a predetermined time frame, the availability of satellites based on their elevation angles at a particular instance of time, and the duration for which each satellite remains accessible to the receiver. This is achieved by establishing a designated Earth surface location as the ground station and setting an elevation mask. Consequently, any satellites that fall below the defined elevation mask at any given point in time will be automatically disregarded. Fig.~\ref{fig1} presents a screenshot depicting the simulated satellite scenario. Refer to this figure, the satellites with a blue line joining itself and the receiver have an elevation angle greater than the elevation mask, indicating that they are available for reception by the receiver. 

\subsection{Receiver Position - Skydel GNSS Simulator}
As another key input to the QuaDRiGa channel generator, we generate the receiver positions in the Earth-centered Earth-fixed (ECEF) coordinate system using the Skydel GNSS simulator. This software offers multiple methods for generating receiver positions, including selecting the starting and ending points on the Skydel GNSS simulator map or importing a keyhole markup language (KML) street map into the software. 

The Skydel GNSS simulator is a software-defined global navigation satellite system (GNSS) simulator that emulates the signals and behavior of various satellite navigation systems, such as GPS, GLONASS, Galileo, BeiDou, and more. It is designed to create realistic and customizable scenarios for testing and evaluating GNSS receivers and systems in a controlled and repeatable environment. To the best of our knowledge, the simulation approach employed by the Skydel GNSS simulator does not adequately replicate the intricate complexities of real-life multipath effects. As a result, the primary utility of the Skydel GNSS simulator lies in generating receiver positions. It is worth mentioning that this software can also generate satellite scenarios. However, given the specific focus of this study which is to efficiently identify available satellites within a defined time frame, we find it more effective to utilize the Matlab Satellite Communications Toolbox for simulating satellite scenarios. 

\section{Challenges and Possible Mistakes/Typos in QuaDRiGa}
In this section, we present a number of technical challenges encountered in the channel modeling using QuaDRiGa to simulate dual mobile space-ground links, followed by a few possible mistakes/typos that we have identified in the source code provided by QuaDRiGa.

\subsection{Challenges in Channel Modeling using QuaDRiGa}
Other than the primary challenge in using the QuaDRiGa channel generator, which is understanding the overall data flow and the methods adopted in the provided source scripts, a number of additional technical challenges are enumerated as follows:

\begin{enumerate}
    \item \textit{Coordinate Transformation}: In the QuaDRiGa reference coordinate system, the origin (0,0,0) corresponds to the position of a chosen receiver at time zero. As a result, the initial position coordinates of all the other receivers and transmitters, if present, need to be adjusted relative to this reference position.

    \item \textit{Antenna Orientation}: To correctly generate channel information, the transmit antennas equipped on satellites should be oriented such that they are pointing toward the receiver at all time. This is accomplished by calculating the elevation and azimuth angle of the satellite and considering the geometric orientation of the antenna defined in QuaDRiGa in conjunction with the reference system for aircraft principal axes.

    \item \textit{Compatibility with State Transition Model}: As the occurrence of a state transition not only depends on the state transition model governed by \eqref{eq1} but also the minimum state duration, it is advisable to sample the terminal trajectory at a fine granularity while comparing the cumulative distance travelled by the receiver with the minimum state duration.
    
     \item \textit{Parameters in Class Track}: The class track in QuaDRiGa is defined to describe the movement of a mobile terminal. While certain parameters within this class lack detailed information, supplementary details for selected parameters are furnished as follows:
    \begin{enumerate}
        \item initial\_position: The initial position of an object should be adjusted relative to the reference position.
        \item positions: The positions of an object should be adjusted relative to its initial position. 
        \item movement\_profile: This parameter stores the distance values in the second row with the corresponding timestamp in the first row. It is used to describe the distance travelled by an object between the current timestamp and the previous one. Consequently, it is updated based on the position at the previous time epoch.  
    \end{enumerate}
\end{enumerate}

\subsection{Possible Mistakes/Typos in Source Code}
In the process of developing the channel modeling tutorial for our target scenario, we identify the following possible mistakes and typos in the source code provided by QuaDRiGa:
\begin{enumerate}
    \item \textit{``get\_channels.m'' in @qd\_layout}: The line 184 in this script serves to determine the current speed of an object based on its movement profile. We believe it should be corrected as follow:
    \begin{multline}
        v = ( mp(2,2:end) - mp(2,1:end-1) ) \\
        ./ ( mp(1,2:end) - mp(1,1:end-1) ).\label{eq15}
    \end{multline}

    \item \textit{``get\_pl.m'' in @qd\_builder}: The line 496 of this script is utilized to calculate atmospheric attenuation. We suspect that there should be a transpose sign followed when invoking the function $qf.interp()$.

    \item \textit{``merging\_cost\_fcn.m'' in private folder of @qd\_builder}: The line 98 in this script aims to calculate the mean square error (MSE) of the delay spread. The original code employs a denominator that encompasses the summation of the target delay spread, thereby exhibiting characteristics more aligned with normalization rather than averaging. Therefore, this line of code is modified as follows:
    \begin{multline}
        mse = 10*\log10( sum( ( ds\_target - ds ).^2 )\\ ./numel(ds))\label{eq16}
    \end{multline}

    \noindent where $ds$ is the name of a parameter storing delay spreads and $numel(ds)$ denotes the number of samples in the parameter $ds$. The revised code serves to calculate the correct MSE value of the delay spread.
\end{enumerate}

\section{Analytical Metrics}
In this section, the method for calculating the received signal power and amplitude, the general approach adopted by QuaDRiGa in generating channel and delay coefficient, the method for analyzing the delay spread, and the technique for computing the Doppler shift are discussed. Among these, the received signal power and amplitude, the multipath delay as well as its distribution, the root mean square (RMS) delay spread, and the Doppler shift collectively serve as the key analytical metrics to validate the accuracy of the generated channel and delay coefficient. 

\subsection{Received Signal Power and Amplitude}

\begin{table}[t]
\caption{Constant Parameters for Satellite Pathloss Model}
\begin{center}
\begin{tabularx}{\linewidth}{>{\centering\arraybackslash}m{1cm}|>{\centering\arraybackslash}X|>{\centering\arraybackslash}X}
\Xhline{1pt}
& \textbf{NTN Urban LOS} & \textbf{NTN Urban NLOS}\\
\Xhline{1pt}
$A$ & 20 & 20.05\\
\hline
$B$ & 32.55 & 54.85\\
\hline
$C$ & 20 & 27.9\\
\hline
$D$ & 0 & -11\\
\Xhline{1pt}
\end{tabularx}
\label{tab2}
\end{center}
\end{table}

Received power serves as an indicator of the pathloss experienced by the signal along its propagation path. In QuaDRiGa, the satellite pathloss model is derived from \cite{b22} which is a combination of the 3GPP-NTN pathloss model and the clutter-loss model. This pathloss model is formulated as follows:
\begin{equation}
\textsf{PL} = A\log_{10}(d_{\text{3d}})+B+C\log_{10}(f_{\text{c}})+D\log_{10}(\alpha)+\textsf{PL}_a\label{eq2}
\end{equation}
\noindent where \textsf{PL} denotes the pathloss of a satellite signal, $d_{\text{3d}}$ denotes the 3D distance between the TX and the RX in meter, $f_{\text{c}}$ denotes the carrier frequency in GHz, $\alpha$ denotes the elevation angle of the TX with respect to the RX in rad, $A$ is a scaling factor that is dependent on the 3D distance between the TX and the RX, $B$ is the reference pathloss at a carrier frequency of 1 GHz and an elevation of 1 rad or 57.3\textdegree, $C$ is a frequency dependent scaling factor, and $D$ is a elevation angle dependent sacling factor. These four constant parameters, $A$, $B$, $C$, and $D$, are scenario specific and are provided in Table \Romannum{2}. $\textsf{PL}_a$ is used to account for the absorption attenuation due to atmospheric gases, which can be neglected when frequencies are below 10 GHz and elevation angles are above 10 degrees \cite{b6,b26}. The channel coefficients are scaled based on the computed pathloss in conjunction with K-factor and shadow fading. 

With the genereted channel coefficient $h$ of dimension $L \times T$, where $L$ denotes the number of paths and $T$ denotes the number of snapshots, the received power of a satellite can be calculated as follows:
\begin{equation}
P^{i}_t = \sum_{l=1}^L |h_{l,t}^{i}|^2\label{eq3}
\end{equation}
\noindent where $P^i_t$ denotes the received power of satellite $i$ at snapshot $t$, $l$ denotes the path index, and $h_{l,t}^{i}$ denotes the channel coefficient of the $l^{th}$ path at snapshot $t$ for satellite $i$. The received signal amplitude of a satellite can be calculated using the generated channel coefficients as follows:
\begin{equation}
A^{i}_t = \sum_{l=1}^L |h_{l,t}^{i}|
\end{equation}

\noindent where $A^i_t$ denotes the received signal amplitude of satellite $i$ at snapshot $t$. One way to validate the propagation scenarios employed by QuaDRiGa is to observe the distribution of the received signal amplitude. When the received signal involves both direct/LOS path and multipaths, the PDF of the signal should follow the Rician distribution as follows:
\begin{equation}
f(a) = \frac{a}{\sigma^2}\text{exp}(-\frac{a^2+z^2}{2\sigma^2})I_0(\frac{az}{\sigma^2})
\end{equation}

\noindent where $a$ denotes the signal amplitude as a random variable, and $\sigma^2$ denotes average multipath power, $z$ is the LOS signal amplitude, and $I_0(\cdot)$ is the modified Bessel function of zeroth order. When the received signal involves multipaths only, the PDF of the signal reduces to the Rayleigh distribution as follows:
\begin{equation}
f(a) = \frac{a}{\sigma^2}\text{exp}(-\frac{a^2}{2\sigma^2}).
\end{equation}

\noindent Both Rayleigh and Rician distribution are employed to fit the histogram of the received signal amplitude in Section \Romannum{6}.

\subsection{Multipath Delay Distribution}
Multipath delay distribution refers the statistical distribution of the multipath delay relative to the direct path. It is a crucial analytical metric for understanding the channel characteristics and has been shown to be mainly subject to the receiver's environment \cite{b17}. In \cite{b20}, the authors conduct a statistical analysis on the GPS multipath in the urban environment. They reveal a notable likelihood of encountering short-delay multipaths compared to those with longer delays in urban areas. This observation will be employed as one of the benchmarks to validate the precision of the channel model we have adopted, by comparing it with the multipath delays in our study. Moreover, we conduct a comparison of the relationship between multipath delay and elevation angle, as well as the multipath delay distribution, with the findings presented in \cite{b17}, which utilizes real GPS measurements collected in a dense urban area, and \cite{b18}, which relies on the real GPS measurements collected in urban areas as discussed in \cite{b35}. There are three distribution functions which are commonly employed to characterize multipath delays: the gamma distribution \cite{b17}, the exponential distribution \cite{b18}, and the Rayleigh distribution \cite{b19}. These distribution functions are employed to fit the multipath delay histogram presented in Section \Romannum{6}.

\subsection{Delay Spread}

\begin{table}[t]
\caption{Parameters for Initial Delay Spread Calculation}
\begin{center}
\begin{tabularx}{\linewidth}{>{\centering\arraybackslash} m{4cm}|>{\centering\arraybackslash}X|>{\centering\arraybackslash}X}
\Xhline{1pt}
\textbf{\textit{\centering Note: $\mu_{\textnormal{DS}}$, $\sigma_{\textnormal{DS}}$, $\alpha_{\textnormal{DS}}$, $\gamma_{\textnormal{DS}}$ ~are calculated using $\mathrm{log}_{10}()$}} & \textit{\centering \textnormal{LOS}} & \textit{\centering \textnormal{NLOS}} \\
\Xhline{1pt}
\centering $\mu_{\textnormal{DS}}$ [dB]& -7.8 & -6.85 \\
\hline
\centering $\sigma_{\textnormal{DS}}$ [dB]& 0.3 & 0.15\\
\hline
\centering $\alpha_{\textnormal{DS}}$ [dB]& 0.5 & N/A\\
\hline
\centering $\gamma_{\textnormal{DS}}$ [dB]& -0.4 & N/A\\
\hline
\centering $d_\lambda$ [m]& 50 & 40\\
\Xhline{1pt}
\end{tabularx}
\label{tab3}
\end{center}
\end{table}

Delay spread describes the temporal spreading or spreading in time of a signal due to multipath propagation in a communication channel. It is another key parameter used to characterize wireless communication channels, particularly in environments with reflections, scattering, and refractions, leading to multiple signal paths arriving at the receiver with different time delays. In QuaDRiGa, for a given carrier frequency, the initial delay spreads for the NTN urban LOS, $\text{DS}_{\text{LOS}}$, and NLOS scenarios, $\text{DS}_{\text{NLOS}}$, are calculated as follows:
\begin{equation}
\text{DS}_{\text{LOS}} = \mu_{\text{DS}} + \gamma_{\text{DS}}  \mathrm{log}_{10}f_{\text{c}} + \alpha_{\text{DS}}  \mathrm{log}_{10}\alpha + \sigma_{DS} X^{\text{DS}}\label{eq4}
\end{equation}
\begin{equation}
\text{DS}_{\text{NLOS}} = \mu_{\text{DS}}  + \sigma_{\text{DS}} X^{\text{DS}}\label{eq5}
\end{equation}

\noindent where $\mu_{\text{DS}}$ denotes the reference value of the delay spread at a carrier frequency of 1 GHz and an elevation of 1 rad or 57.3\textdegree, $\sigma_{\text{DS}}$ denotes the standard deviation of the delay spread at the reference value, $\gamma_{\text{DS}}$ denotes a frequency dependent scaling factor, $\alpha_{\text{DS}}$ denotes an elevation dependent scaling factor, and $X^{\text{DS}}$ denotes a random variable that follows a spatially correlated normal distribution with zero mean, unit variance, and an autocorrelation function which is a mixture of a Gaussian and an exponential autocorrelation function:
\begin{equation}
\rho(d) = \left\{
\begin{array}{ll}
    \exp(-d^2/d_\lambda^2), & \text{if } d < d_\lambda \\
    \exp(-d/d_\lambda), & \text{if } d \geq d_\lambda 
\end{array}
\right.
\label{eq6}
\end{equation}
\noindent where $d_\lambda$ represents the decorrelation distance and $d$ is the distance between two mobile terminals (MTs), in our case, the TX (e.g., satellite) and the RX (e.g., ground user). Due to the large distance between the TX and the RX in our simulation, the condition $d \geq d_\lambda$ is always satisfied. The parameters used for the calculation of the initial delay spread are provided in the source code of QuaDRiGa and are summarized in Table \Romannum{3}. To consider the relationships among large-scale parameters including delay spread and others such as Ricean K-factor, shadow fading, azimuth spread of departure, azimuth spread of arrival, elevation spread of departure, elevation spread of arrival, and cross-polarization ratio, an inter-parameter correlation model is employed. The detailed information on generating the spatially correlated large-scale parameters can be found in \cite{b6}. The spatially correlated large-scale parameters are employed to generate initial delays, cluster powers, and drifting angle of arrivals. These parameters are then combined with the defined antenna patterns and array geometries to generate channel and delay coefficients. 

The resulting channel and delay coefficients are validated by examining the RMS delay spread that is computed by
\begin{equation}
\text{DS}_{\text{rms}} = \sqrt{\frac{1}{P}\sum_{l=1}^L P_l\tau_l^2 - (\frac{1}{P}\sum_{l=1}^L P_l\tau_l)^2}\label{eq7}
\end{equation}
\noindent where $P$ denotes the sum of all path power $P_l$, and $\tau_l$ denotes the $l^{th}$ path delay.

\subsection{Doppler Spectrum}

\begin{figure}[t]
\centerline{\resizebox{\columnwidth}{!}{\includegraphics{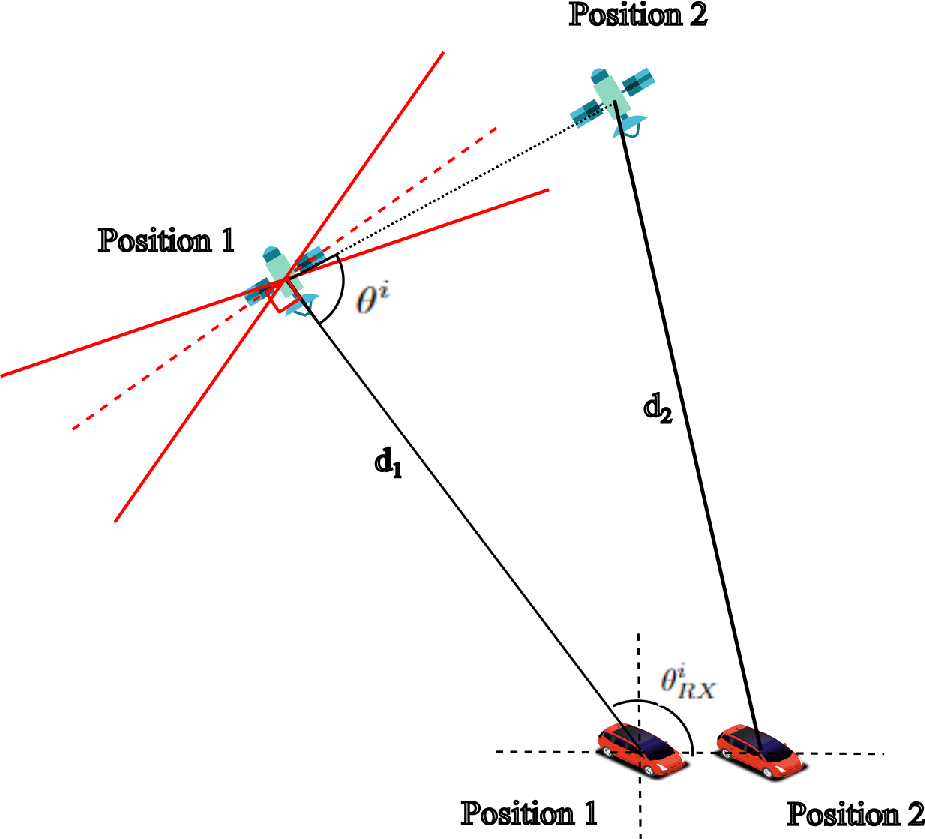}}}
\caption{Doppler angle illustration (red solid line: upper and lower bounds of the Doppler angle; red dash line: the 90\textdegree ~Doppler angle).}
\label{fig9}
\end{figure}

Doppler spectrum refers to the distribution of the signal power as a function of the Doppler shift. It is commonly used in wireless communications and navigation systems to analyze the frequency shift of signals caused by the relative motion between the transmitter and the receiver. In this work, the Doppler spectrum is found by calculating the frequency response of the channel, followed by an inverse Fast Fourier Transform (iFFT). The detailed steps can be found in \cite{b6}. 

To validate the Doppler effect, the most prominent Doppler shift depicted in the Doppler spectrum is compared with the estimated Doppler shift for each satellite, which is given as follows:
\begin{equation}
\delta f^i = \frac{f_\text{c}}{c}\delta v^i \label{eq8}
\end{equation}

\noindent where $\delta f^i$ denotes the Doppler shift for satellite $i$, $c$ is the speed of light, and $\delta v^i$ denotes the radial relative velocity between satellite $i$ and the receiver, which can be calculated by
\begin{equation}
\delta v^i = v^i\cos{\theta^i} + v_{\text{RX}}\cos{\theta_{\text{RX}}^i} \label{eq9}
\end{equation}

\noindent where $v^i$ denotes the speed of satellite $i$, $\theta^i$ denotes the angle between the movement direction of satellite $i$ and the LOS between satellite $i$ and the receiver, which is commonly referred to as the Doppler angle of satellite $i$, $v_{\text{RX}}$ denotes the speed of the receiver, and $\theta_{\text{RX}}^i$ denotes the angle between the direction of receiver's motion and the LOS between satellite $i$ and the receiver, or the Doppler angle of the receiver. A visual illustration of the Doppler angles for the satellite and the receiver is depicted in Fig.~\ref{fig9}. It should be noted that the sign of radial relative velocity is determined by the motion trajectories of the satellite and the receiver. Refer to Fig.~\ref{fig9}, suppose the distance between the satellite and the receiver at position 1, denoted as $d_1$, is greater than the distance at position 2, denoted as $d_2$, meaning that the satellite is approaching the receiver, then the sign of radial relative velocity should be positive. Based on the fact that the maximum radial relative velocity between a stationary receiver and a GPS satellite is about 929 m/s \cite{b21}, the upper and lower bounds of the Doppler angle at varied satellite speeds can be determined. Therefore, we conduct a sanity check on the celestial mechanics by verifying the correctness of the Doppler angles for the simulated GPS satellites. According to \eqref{eq9}, a satellite with a Doppler angle of 90\textdegree~would yield no radial velocity. The upper bound and lower bound of the Doppler angle along with the 90\textdegree ~Doppler angle are also shown in Fig.~\ref{fig9}.

\section{Simulation Setup}
In this study, a tutorial for the channel modeling of an urban dual-mobility scenario for the space-ground links is developed. The simulation is carried out for a duration of 1 second. The elevation mask is set to 15 degrees. The carrier frequency of the transmitted signal (GPS L1) is 1,575.42 MHz. Since satellites are moving at a great speed (e.g., 3,000 m/s), the channel update rate needs to be sufficiently large in order to satisfy the sampling theorem and accurately capture the Doppler effect. We choose to use a sample density of 5 samples per wavelength for the dual-mobility scenario. This is the default value recommended by QuaDRiGa to correctly capture the Doppler characteristics. As the longest distance travelled by a satellite in this work is around 3,500 meters, we can calculate the minimum channel update rate, $R_{\text{ch}}$, required for the simulation as follows:
\begin{equation}
R_{\text{ch}} = 1 \times 3500 \times 5 \times \frac{f_{\text{c}}}{c} = 91.963 ~\mathrm{kHz}.\label{eq10}
\end{equation}

\begin{figure}[t]
\centerline{\resizebox{\columnwidth}{!}{\includegraphics{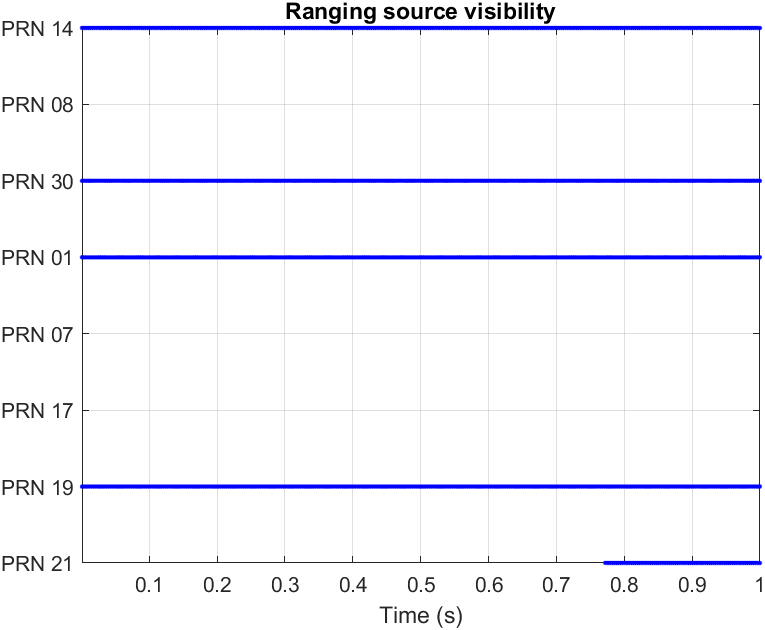}}}
\caption{Satellite visibility plot (blue line indicates a LOS scenario).}
\label{fig3}
\end{figure}

\noindent To simplify the computation, we opt for a channel update rate of 100 kHz. A propagation scenario, either NTN urban LOS or NTN urban NLOS, is assigned to each satellite every 1 ms. In this simulation, the speed of the RX (e.g., vehicle) is 50 km/h, thence the distance travelled by the vehicle for one second is about 14 meters. It can be observed that a transition between states can occur within a simulation time of 1 second, as evident from the minimum state duration values presented in Table \Romannum{1}. All satellite antennas are featured with a parabolic dish antenna with right-hand circular polarization (RHCP), an aperture radius of 3 meters, and a gain of 40 dBi. A 0 dBm transmit power is applied to the satellite. A patch antenna pointing to the sky is equipped on the receiver. It is worth noting that we perform the entire simulation using Matlab, however, the channel modeling by QuaDRiGa can also be performed in Octave where graphics processing unit (GPU) can be utilized to accelerate the simulation. In the calculation of the Doppler spectrum for the satellites, a Doppler analysis window size of 1 second is chosen to optimize frequency resolution. Following the typical parameter selection for GPS L1 C/A code, a channel bandwidth of 2 MHz  is considered\cite{b37,b38}. To render a considerably good performance, a Fast Fourier Transform (FFT) point of 1024 is adopted. The key simulation parameters are summarized in Table \Romannum{4}.

\begin{table}[t]
\caption{Key Simulation Parameters}
\begin{center}
\begin{tabularx}{\linewidth}{>{\centering\arraybackslash}X|>{\centering\arraybackslash}X}
\Xhline{1pt}
\textbf{Parameter}     &\textbf{Value}\\
\Xhline{1pt}
Simulation duration        &1 s\\
\hline
Elevation mask        &15 deg\\
\hline
Carrier frequency        &1,575.42 MHz\\
\hline
Sample density      &5 samples/wavelength\\
\hline
Channel update rate ($R_{\text{ch}}$)      &100 kHz\\
\hline
Vehicle (receiver) speed        &50 km/h\\
\hline
Transmit antenna gain        &40 dBi\\
\hline
Transmit power        &0 dBm\\
\hline
Channel bandwidth        &2 MHz\\
\hline
FFT-point        &1024\\
\Xhline{1pt}
\end{tabularx}
\label{tab5}
\end{center}
\end{table}

\section{Simulation Results}

\begin{table}[t]
\caption{Starting and Ending Elevation Angle for Satellites}
\begin{center}
\begin{tabularx}{\linewidth}{>{\centering\arraybackslash}X|>{\centering\arraybackslash}X|>{\centering\arraybackslash}X}
\Xhline{1pt}
\textbf{Satellite ID} & \textbf{$\alpha_i$ [deg]} & \textbf{$\alpha_f$ [deg]}\\
\Xhline{1pt}
PRN 14	&72.6560	& 72.6638\\
\hline
PRN 08	&15.7490	& 15.7431\\
\hline
PRN 30	&47.7190	& 47.7120\\
\hline
PRN 01	&63.4044	& 63.4049\\
\hline
PRN 07	&21.6368	& 21.6290\\
\hline
PRN 17	&47.8612	& 47.8676\\
\hline
PRN 19	&17.7293	& 17.4352\\
\hline
PRN 21	&45.7599	& 45.7546\\
\Xhline{1pt}
\end{tabularx}
\label{tab4}
\end{center}
\end{table}

Under the simulation setup elucidated in the preceding section, the simulation time required by QuaDRiGa for channel modeling turns out to be 36 minutes. A total of about 600 MB (in binary) data for the network layout and the channel information is collected for post-processing. The satellite visibility over a simulation period of 1 second is depicted in Fig.~\ref{fig3}. All satellites shown in this figure are considered available to the RX. Notably, we can observe that the LOS condition is eventually established for the satellite with the pseudo random noise (PRN) code 21 due to the state transition model. The frequency of state transitions may increase if the simulation is prolonged over an extended time frame. The starting and ending elevation angles, $\alpha_i$ and $\alpha_f$, of each satellite are summarized in Table \Romannum{5}. To validate the generated channel and delay coefficient, the received power, multipath delay distribution, delay spread, and Doppler spectrum are analyzed in this section. It is worth noting that we have the knowledge of the positions for both satellites and the receiver, sampled at a rate optimized by QuaDRiGa for efficient channel modeling. However, it is important to mention that this sampling rate may not align with the channel update rate utilized for generating channel and delay coefficients. As a result, the position data for satellites and the receiver corresponding to each generated path delay might not be accessible.

\subsection{Received Signal Power and Amplitude}
\begin{figure*}[t]
\centering
\begin{tabularx}{\textwidth}{X X X X}
\includegraphics[width=\linewidth]{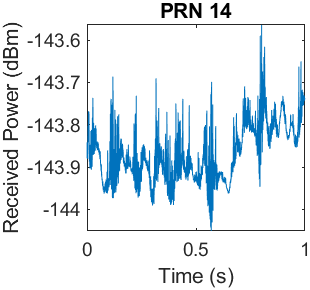}&
\includegraphics[width=\linewidth]{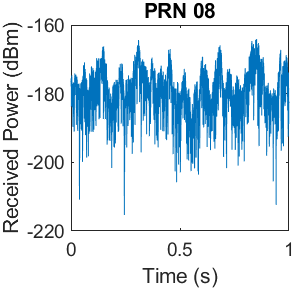}&
\includegraphics[width=\linewidth]{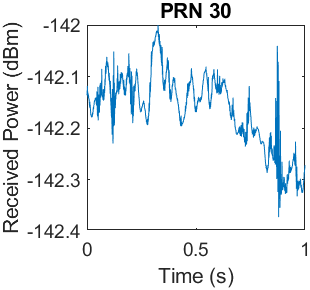}&
\includegraphics[width=\linewidth]{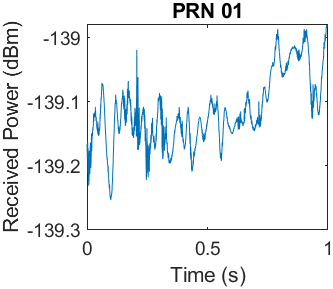}\\
\includegraphics[width=\linewidth]{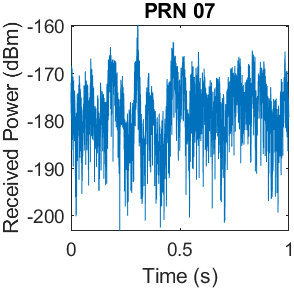}&
\includegraphics[width=\linewidth]{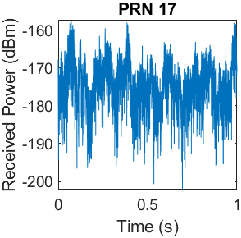}&
\includegraphics[width=\linewidth]{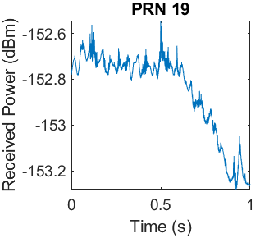}&
\includegraphics[width=\linewidth]{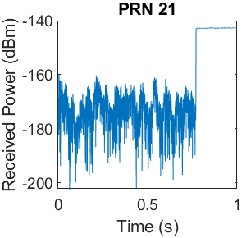}
\end{tabularx}
\caption{Received power for GPS satellites.}
\label{fig4}
\end{figure*}

The received power of the simulated satellites is shown in Fig.~\ref{fig4}. We can observe from this figure that when the LOS condition is established between the satellite and the receiver, the received power exhibits larger amplitude and reduced noise. This observation is logical since the received power of the direct/LOS path is typically much higher than that of the multipath. Using \eqref{eq2} and the constant parameters for the satellite pathloss model provided in Table \Romannum{2}, we can estimate the pathloss for each satellite. For instance, the propagation scenario for satellite PRN 21 is urban-NLOS at the start of the simulation. At this moment, the 3D distance $d_{\text{3d}}$ between satellite PRN 21 and the receiver is roughly found to be 21,954.8287 km and the elevation angle is 45.7599 degrees. The pathloss of satellite PRN 21 can then be computed as follows:
\begin{multline}
    \textsf{PL} = 20.05\log_{10}(21{,}954{,}828.7)+54.85\\
    +27.9\log_{10}(1.57542)-11\log_{10}(45.7599\times\frac{\pi}{180})\\
    = 208.63 ~\mathrm{dB}.
\end{multline}

\noindent Knowing that the transmit power is 0 dBm and the transmit antenna gain is 40 dBi, we can roughly determine the received power for PRN 21 at the start of the simulation, which is -168.63 dBm. This matches with the received power for PRN 21 shown in Fig.~\ref{fig4}. A similar analysis is carried out for an urban-LOS case, in particular, PRN 14 at the start of the simulation. At this moment, we determine that the 3D distance $d_{\text{3d}}$ between the satellite and the receiver is 20,408.4785 km. The pathloss for this satellite is then computed as follows: 
\begin{multline}
    \textsf{PL} = 20\log_{10}(20{,}408{,}478.5)+32.55+20\log_{10}(1.57542)\\
    = 182.69 ~\mathrm{dB}.
\end{multline}
\noindent Following the same analysis, the received power is computed to be -142.69 dBm. This again roughly matches with the received power for PRN 14 as shown in Fig.~\ref{fig4}. 

\begin{figure*}[t]
\centering
\begin{tabularx}{\textwidth}{X X X X}
\includegraphics[width=\linewidth]{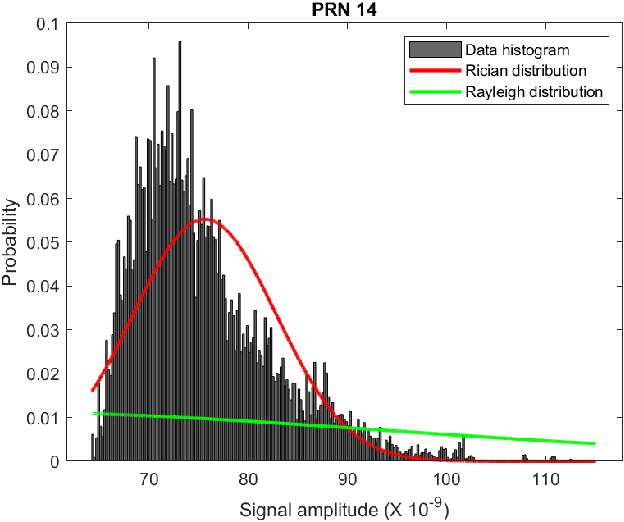}&
\includegraphics[width=\linewidth]{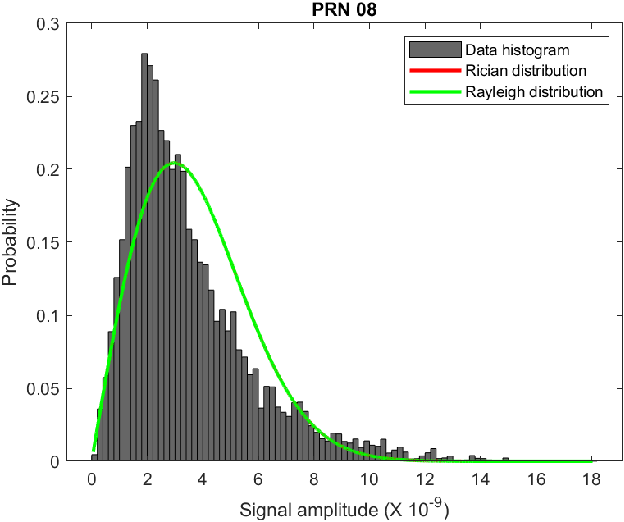}&
\includegraphics[width=\linewidth]{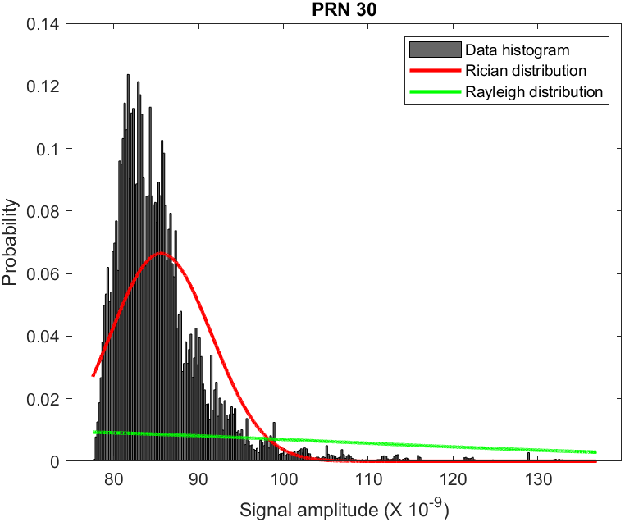}&
\includegraphics[width=\linewidth]{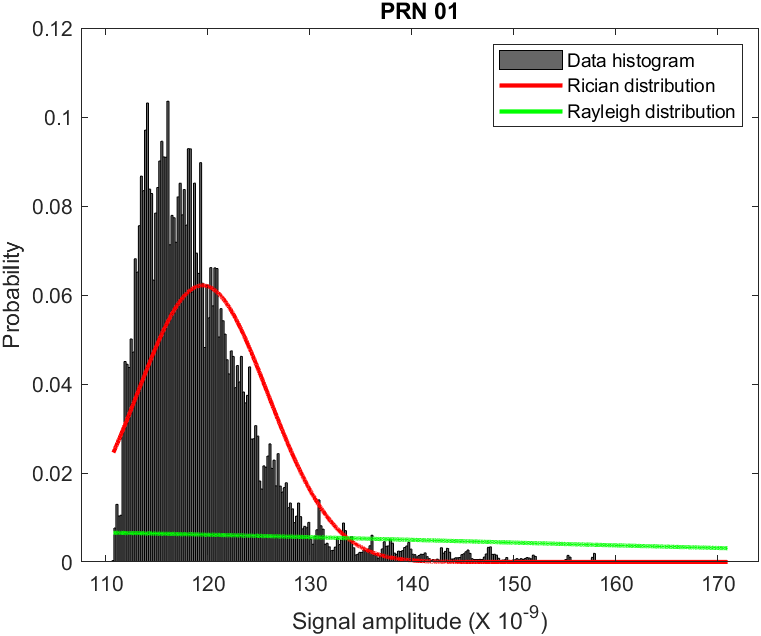}\\
\includegraphics[width=\linewidth]{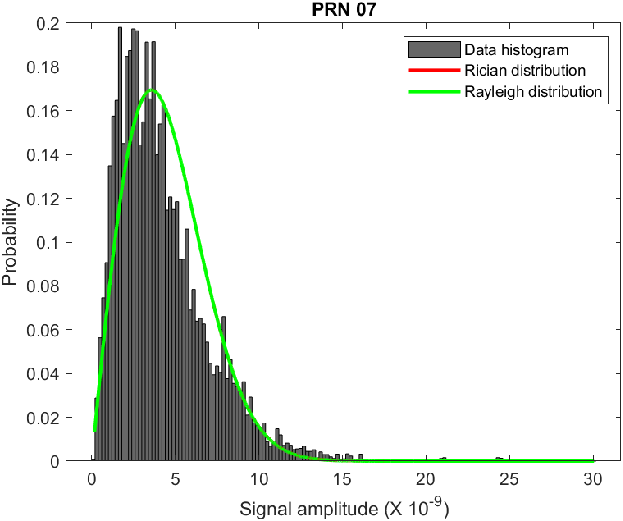}&
\includegraphics[width=\linewidth]{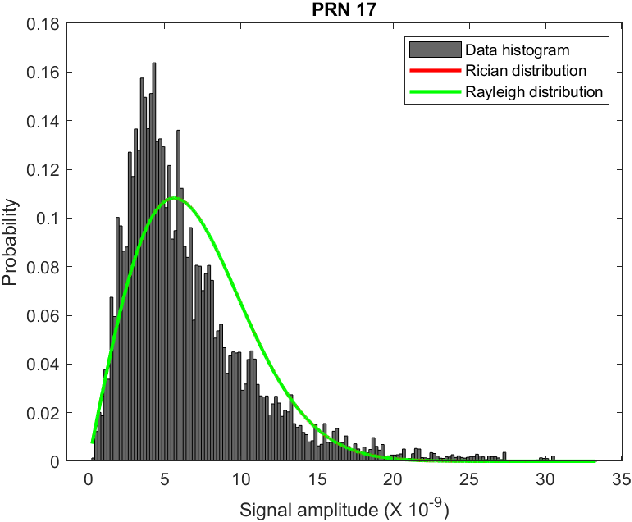}&
\includegraphics[width=\linewidth]{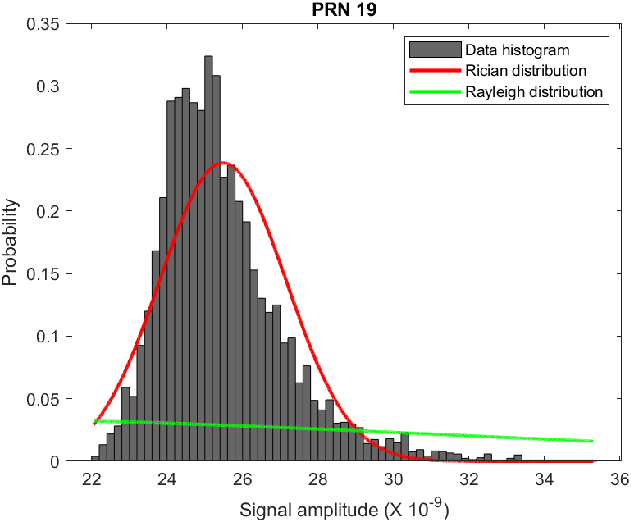}&
\includegraphics[width=\linewidth]{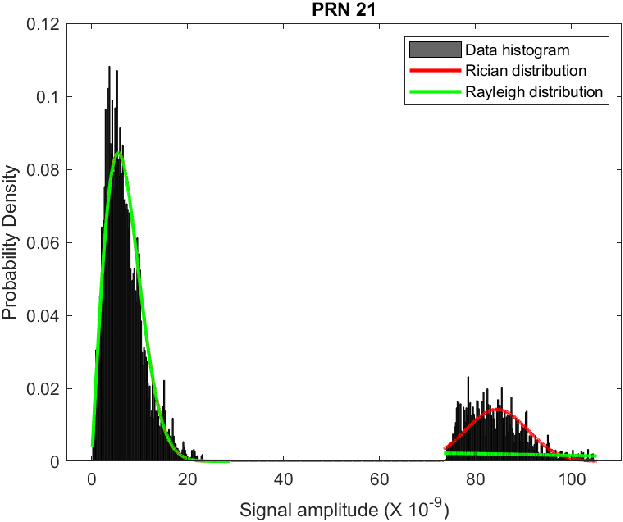}
\end{tabularx}
\caption{Histogram of the received signal amplitude with fitted Rayleigh and Rician distribution.}
\label{fig11}
\end{figure*}

Fig.~\ref{fig11} presents the histogram of the received signal amplitude for simulated satellites, alongside the fitted Rayleigh and Rician distributions. This illustration reveals that the Rician distribution converges to the Rayleigh distribution when the received signal is solely influenced by multipath. However, in the presence of a LOS path in the received signal, the Rayleigh distribution is not a suitable fit for the histogram of the received signal amplitude. The histogram of the received signal amplitude for PRN 21 exhibits two distinct distributions. This is attributed to the initial NLOS propagation scenario for PRN 21, which later transitions to a LOS scenario. To validate the propagation scenarios for this satellite, we divide the corresponding received signal amplitude into two datasets where the first one contains low amplitude values, corresponding to the NLOS scenario, and the second one contains high amplitude values, corresponding to the LOS scenario. Both groups are fitted using Rayleigh and Rician distributions. The first dataset shows the Rician fit overlapped with the Rayleigh fit, while only the Rician distribution can be used to fit the second dataset. In summary, the amplitude distribution for all simulated satellites aligns with theoretical predictions, demonstrating the accuracy of the propagation scenarios employed by QuaDRiGa channel generator.

\subsection{Multipath Delay Distribution}
\begin{figure*}[t]
\centering
\begin{tabularx}{\textwidth}{X X}
\includegraphics[width=\linewidth]{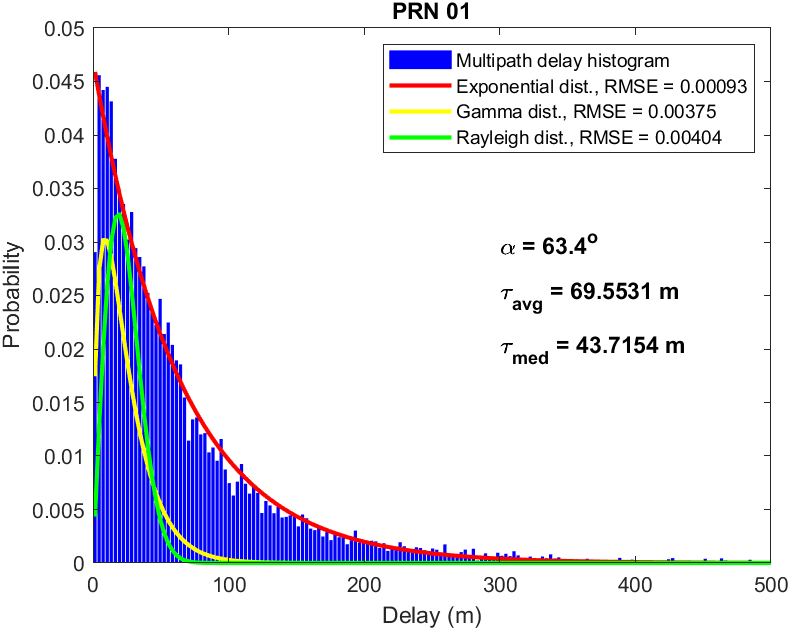}&
\includegraphics[width=\linewidth]{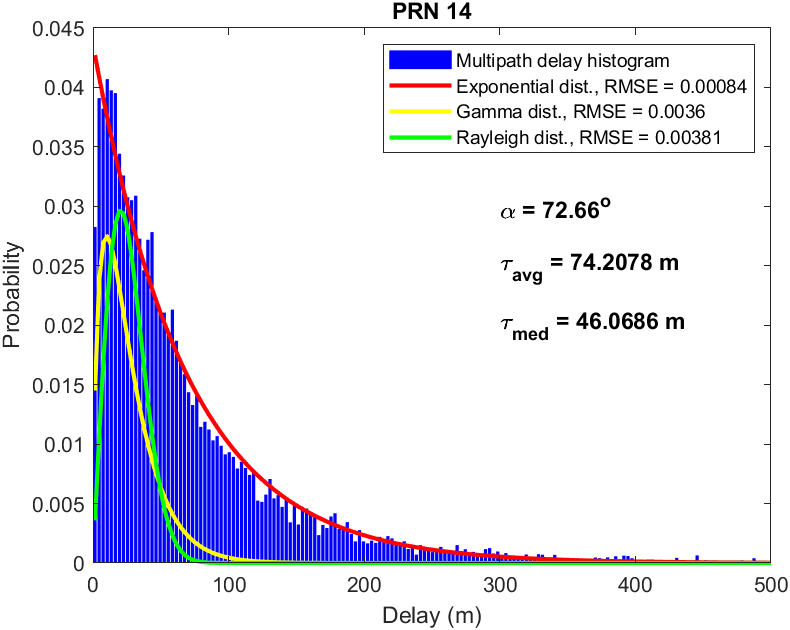}\\
\includegraphics[width=\linewidth]{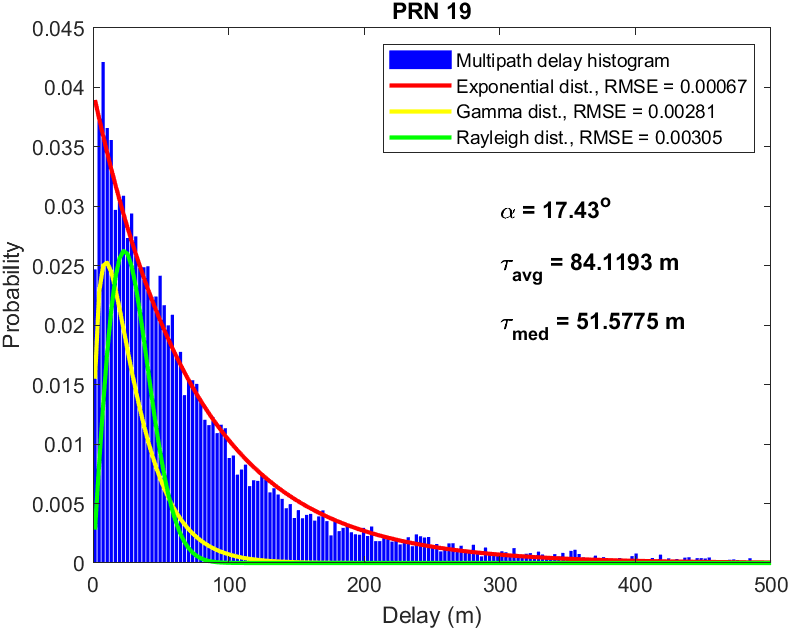}&
\includegraphics[width=\linewidth]{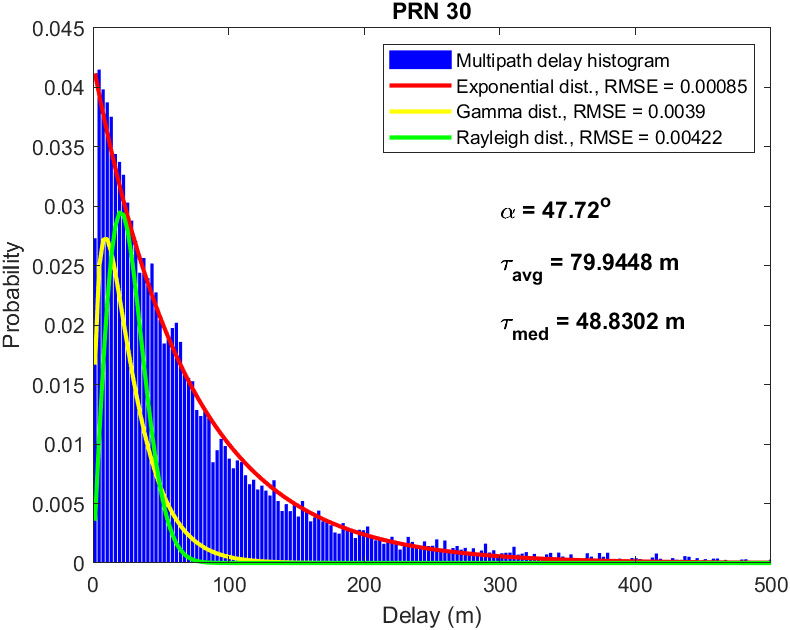}
\end{tabularx}
\caption{Multipath delay histogram for LOS satellites PRN 01, PRN 14, PRN 19, PRN 30.}
\label{fig5}
\end{figure*}

Due to different sampling rates adopted by QuaDRiGa in position interpolation and delay generation, the positions for satellites and the receiver associated with each generated path delay are unavailable. In other words, the delay of the direct path in the NLOS scenarios is unknown to us. Therefore, we will be analyzing the multipath delay distribution for the LOS satellites: PRN 01, PRN 14, PRN 19, and PRN 30. Three candidate distributions mentioned in Section \Romannum{3} are used to fit the histogram for each satellite in order to find out which distribution is the best for multipath delay in the urban scenario considered by QuaDRiGa channel generator. Based on the least squares criteria, the regressive curve fitting method is used to determine the optimal coefficients for all possible distribution functions. The histogram and the corresponding root mean square error (RMSE) for each distribution curve fitting, along with the elevation angle $\alpha$, the average multipath delay, $\tau_{\text{avg}}$, and the median multipath delay $\tau_{\text{med}}$ for each satellite are depicted in Fig.~\ref{fig5}. From this figure, the following information can be observed:

\begin{enumerate}
    \item The average multipath delay $\tau_{\text{avg}}$ and the median multipath delay $\tau_{\text{med}}$ decrease as the elevation angle increase in general. This finding is consistent with the observation made in \cite{b17} in general. However, there can be exceptions to this trend, particularly at high elevation angles. This is because a higher elevation angle may not necessarily correspond to the shortest distance between the satellite and the receiver. 
    \item The exponential distribution is the best fit among the three considered distributions as it has the lowest RMSE value. This result matches with the multipath delay distribution proposed in \cite{b18}, where actual GPS measurements collected from urban areas are utilized. The dominance of the exponential distribution remains consistent across all elevation angles, indicating that the multipath delay is primarily influenced by the receiver's surrounding environment. This also corresponds with the observation made in \cite{b17}.
    \item With QuaDRiGa channel models for the NTN-urban scenario, there is a notably greater likelihood of encountering short-delay multipaths compared to those with longer delays. This observation is in accordance with the conclusion presented in \cite{b20}. 
\end{enumerate}

\subsection{Delay Spread}
\begin{figure}[h]
\centering
\centering\includegraphics[width=\linewidth]{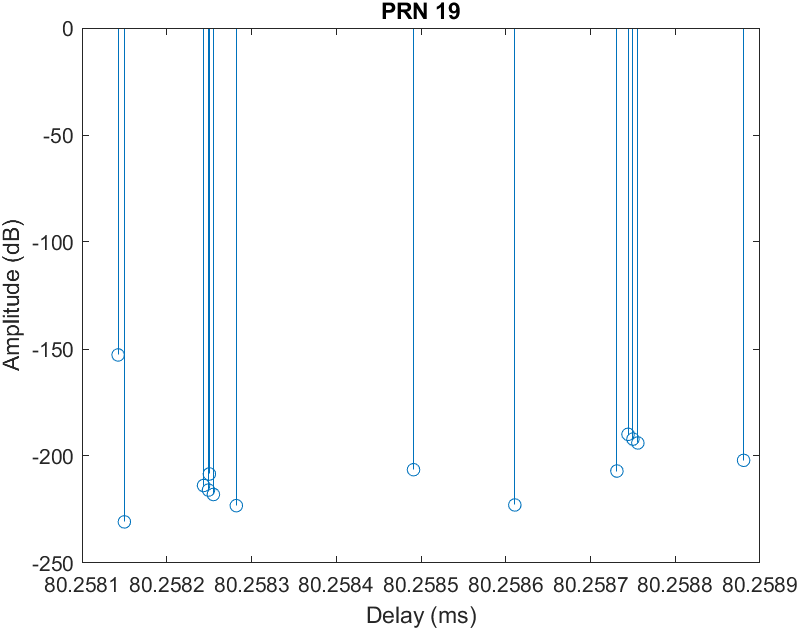}
\caption{Delay spread for PRN 19 at the initial epoch.}
\label{fig10}
\end{figure}

\begin{figure*}[t]
\centering
\centering
\begin{tabularx}{\textwidth}{X X X X}
\includegraphics[width=\linewidth]{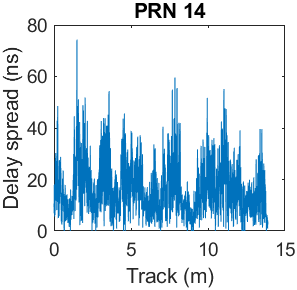}&
\includegraphics[width=\linewidth]{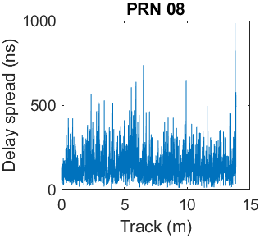}&
\includegraphics[width=\linewidth]{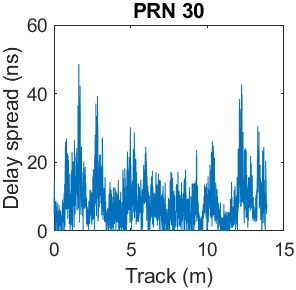}&
\includegraphics[width=\linewidth]{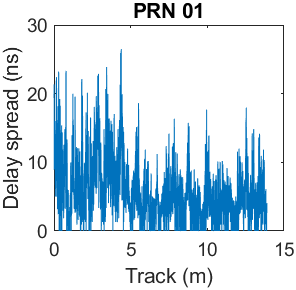}\\
\includegraphics[width=\linewidth]{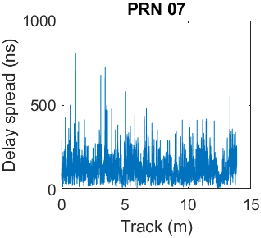}&
\includegraphics[width=\linewidth]{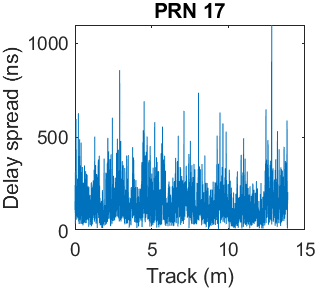}&
\includegraphics[width=\linewidth]{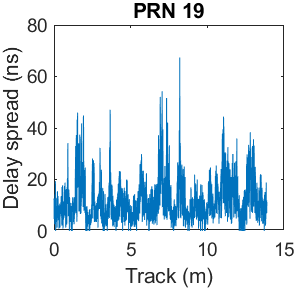}&
\includegraphics[width=\linewidth]{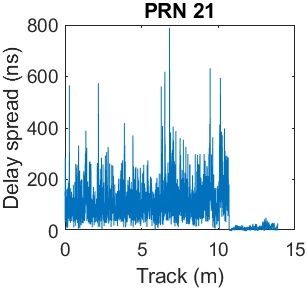}\\
\end{tabularx}
\caption{RMS delay spread for the simulated GPS satellites.}
\label{fig6}
\end{figure*}

Fig.~\ref{fig10} illustrates the delay spread at the initial epoch for satellite PRN 19, in which a LOS condition is established between itself and the receiver. This plot highlights that the signal along the direct path exhibits the most robust signal amplitude. Furthermore, it is noticeable that a multipath signal characterized by a shorter delay does not necessarily align with a stronger amplitude. This demonstrates a successful simulation of the occurrence of constructive and destructive interference which can occur in real life. The RMS delay spread of the simulated satellites is shown in Fig.~\ref{fig6}. The x-axis in the delay spread plots denotes the distance travelled by the RX. Using \eqref{eq4}, \eqref{eq5} and the extracted parameters given in Table \Romannum{1}, the reasonable range of the delay spread of each satellite can be calculated. For example, the delay spread of satellite PRN 14 in the urban scenario with an elevation of 72.6560\textdegree ~can be calculated by substituting $\mu_{\text{DS}}=-7.8$, $\sigma_{\text{DS}}=0.3$, $\alpha_{\text{DS}}=0.5$, and $\gamma_{\text{DS}}=-0.4$ into \eqref{eq4}, resulting
\begin{equation}
\text{DS}_{\text{LOS,urban}}^{\text{PRN 14}} = -7.8274 + 0.3X^{\text{DS}}.
\end{equation}
Since $X^{\text{DS}}$ is a random variable that follows a normal distribution with zero mean and unit variance, we could consider $X^{\text{DS}}$ to be any value in between $-3\sigma$ and $+3\sigma$ in estimating the range of the delay spread. Therefore, the range for $\text{DS}_{\text{LOS,urban}}^{\text{PRN 14}}$ is found to be between 1.8733 ns and 118.1952 ns, which is consistent with the delay spread for PRN 14 shown in Fig.~\ref{fig6}. As an example for the NLOS case, the delay spread of PRN 08 is found using \eqref{eq5} as follows:
\begin{equation}
\text{DS}_{\text{NLOS,urban}}^{\text{PRN 08}} = -6.85 + 0.15X^{\text{DS}}.
\end{equation}

\noindent By the same analysis, the delay spread for PRN 08 falls into the range between 50.1187 ns and 398.1072 ns. We notice that in this case, the calculated range of delay spread does not match perfectly with the delay spread shown in Fig.~\ref{fig6}. This is likely due to a significant Doppler shift of this signal (see Fig.~\ref{fig7}), rendering a prominent time varying channel for the link between PRN 08 and the receiver. Since we assume a nominal carrier frequency when applying \eqref{eq4} and \eqref{eq5} to estimate the initial delay spread, it is reasonable to anticipate the calculated range for delay spread to be more accurate for the ones with less Doppler shift (i.e., frequency shift).

\subsection{Doppler Spectrum}
\begin{figure*}[t]
\centering
\begin{tabularx}{\textwidth}{X X X X}
\includegraphics[width=\linewidth]{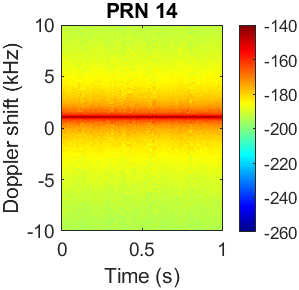}&
\includegraphics[width=\linewidth]{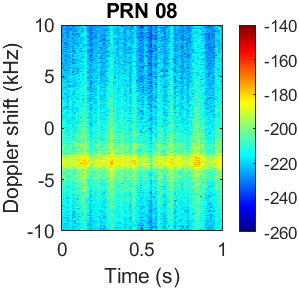}&
\includegraphics[width=\linewidth]{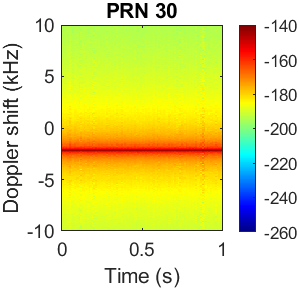}&
\includegraphics[width=\linewidth]{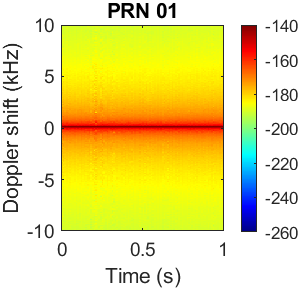}\\
\includegraphics[width=\linewidth]{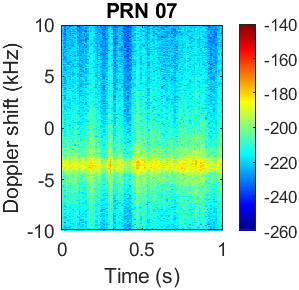}&
\includegraphics[width=\linewidth]{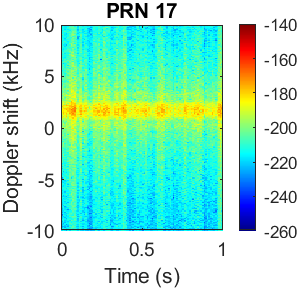}&
\includegraphics[width=\linewidth]{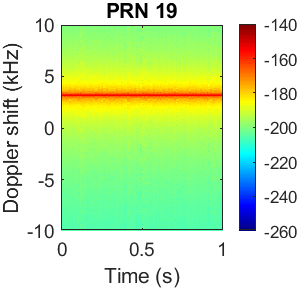}&
\includegraphics[width=\linewidth]{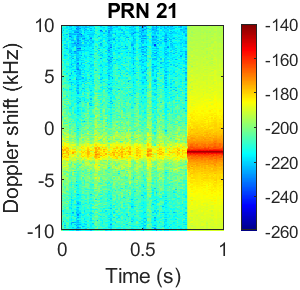}
\end{tabularx}
\caption{Doppler spectrum for the simulated GPS satellites.}
\label{fig7}
\end{figure*}

The Doppler spectrum of the simulated satellites is shown in Fig.~\ref{fig7}. From Fig.~\ref{fig7}, it is evident to see that the Doppler shifts of all satellites lie in between -5 kHz and +5 kHz. This observation aligns with the expectation of the Doppler shift for a space-ground link that involves a stationary or a slow moving receiver \cite{b16,b21}. Additionally, we can observe that the intensity of the Doppler spectrum in the LOS scenario appears to be stronger compared to the NLOS scenario. Furthermore, the Doppler shift in the NLOS scenario exhibits more noise when contrasted with the LOS scenario. This aligns with the anticipated behavior as the power of MPCs is considerably lower than that of the direct path in general. A specific instance illustrating both of these characteristics can be observed in the case of PRN 21. To validate the accuracy of the Doppler spectrum generated using the channel and delay coefficient, we compare the strongest Doppler shift in the spectrum with the one calculated using \eqref{eq8}. For instance, by knowing the positions of both satellite PRN 21 and the receiver at two consecutive epochs with known time interval, the radial relative velocity for this specific satellite-receiver pair is estimated to be -434.5713 m/s. Based on this information, the Doppler shift for satellite PRN 21 can be calculated as follows:
\begin{equation}
\delta f = \frac{1{,}572.42 \times 10^6}{299{,}792{,}458} \times -434.5713 = -2{,}283.69 ~\mathrm{Hz}.
\end{equation}

\begin{figure}[t]
\centering
\centering\includegraphics[width=\linewidth]{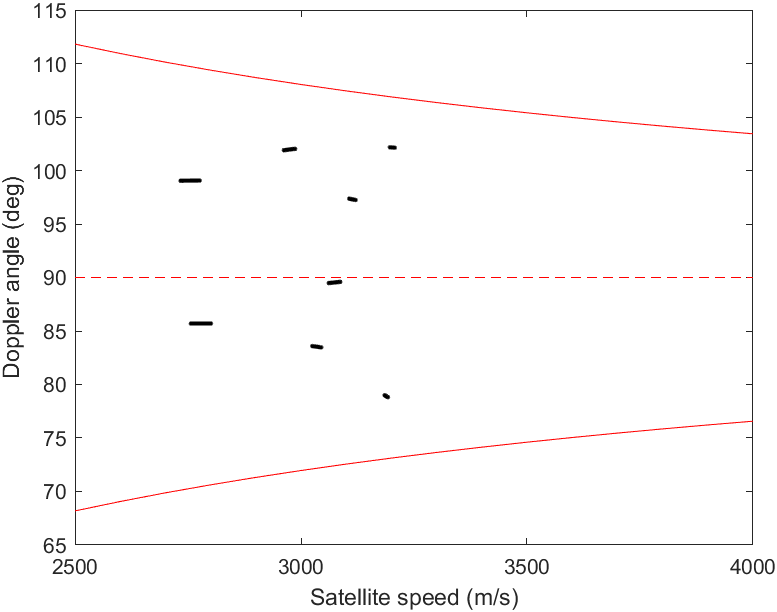}
\caption{Doppler angles for the simulated GPS satellites (black dots: simulated GPS satellites; red solid lines: upper and lower bound of the Doppler angle for GPS satellites; red dash line: the Doppler angle at which zero radial velocity is contributed by the satellite).}
\label{fig8}
\end{figure}

\noindent This roughly aligns with strongest Doppler shift of satellite PRN 21 in the Doppler spectrum as shown in Fig.~\ref{fig7}. By assuming the receiver is stationary, the plot of the Doppler angle for the simulated satellites along with the upper and lower bounds of the Doppler angle for different satellite speeds is shown in Fig.~\ref{fig8}. As we can observe from this figure, the Doppler angle bound approaches 90 degrees as the satellite speed increases. Furthermore, it is evident that the Doppler angles of all the simulated satellites fall within the specified upper and lower bounds, demonstrating the fidelity of celestial mechanics involved in the system model.

\section{Conclusion}
In this paper, we have developed a scalable tutorial that can generate realistic channel and delay coefficients for dual mobile space-ground links. This is achieved by utilizing the Matlab Satellite Communications Toolbox, the Skydel GNSS simulator, and the QuaDRiGa channel generator. Although the QuaDRiGa channel generator is complex, essentially, users must provide three key inputs: the trajectory of the terminal along with specified signal propagation scenarios, a network layout, and details of antenna parameters. For simulating realistic signal quality conditions in different propagation scenarios, recognized models for LOS probability and state transition are utilized. QuaDRiGa then automates the remaining channel modeling processes.

The statistical nature of QuaDRiGa eliminates the need for high-resolution maps and surface parameters, enabling fast deployment while maintaining sufficient accuracy. We validate the generated channel and delay coefficients by analyzing the received signal power and amplitude, multipath delay distribution, delay spread and Doppler spectrum. With the intention of functioning as a tutorial, we highlight several technical obstacles that users could encounter while utilizing QuaDRiGa, particularly when simulating the dual mobile space-ground connections. Several possible errors and typographical errors present in the source code supplied by QuaDRiGa are identified to enhance the reliability of the open source.

As the propagation scenario files provided by QuaDRiGa can be utilized for any NTN scenario, we anticipate the applicability of this tutorial to extend to dual mobile stratospheric-ground links and air-ground links. This versatility becomes valuable for integrated signal processing within VHetNets. By convolving the generated channel coefficients with any type of signals, such as satellite and HAPS, and incorporating different delays to the same signal, we can generate pseudo signals in a statistically reliable multipath propagation environment.

%

\begin{IEEEbiography}[{\includegraphics[width=1in,height=1.25in,clip,keepaspectratio]{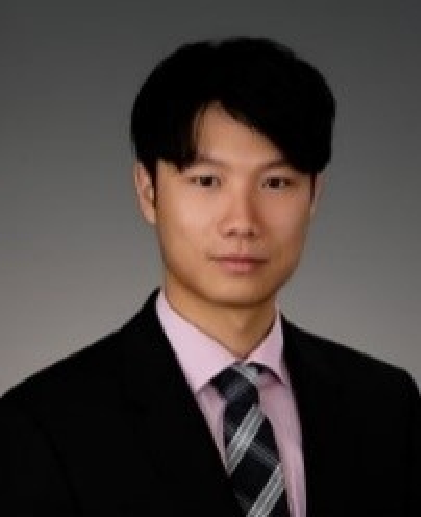}}]{Hongzhao Zheng}
(Member, IEEE) received the B. Eng. (Hons.) degree in engineering physics from the Carleton University, Ottawa, ON, Canada, in 2019. He is currently a PhD student at Carleton University. His research interest is urban positioning using sensor-enabled heterogeneous wireless infrastructure. 

Mr. Zheng received the best paper award at IEEE WiSEE 2022.
\end{IEEEbiography}

\begin{IEEEbiography}[{\includegraphics[width=1in,height=1.25in,clip,keepaspectratio]{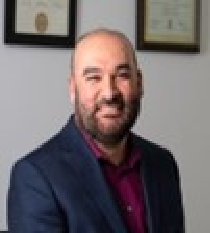}}]{Mohamed Atia}
(Senior Member, IEEE) received the B.S. and M.Sc. degrees in computer systems from Ain Shams University, Cairo, Egypt, in 2000 and 2006, respectively, and the Ph.D. degree in electrical and computer engineering from Queen’s University, Kingston, ON, Canada, in 2013. He is currently an Associate Professor with the Department of Systems and Computer Engineering, Carleton University. He is also the Founder and the Director of the Embedded and Multi-Sensory Systems Laboratory (EMSLab), Carleton University. His research interests include sensor fusion, navigation systems, artificial intelligence, and robotics.

Dr. Atia received the best paper award at IEEE WiSEE 2022.
\end{IEEEbiography}

\begin{IEEEbiography}[{\includegraphics[width=1in,height=1.25in,clip,keepaspectratio]{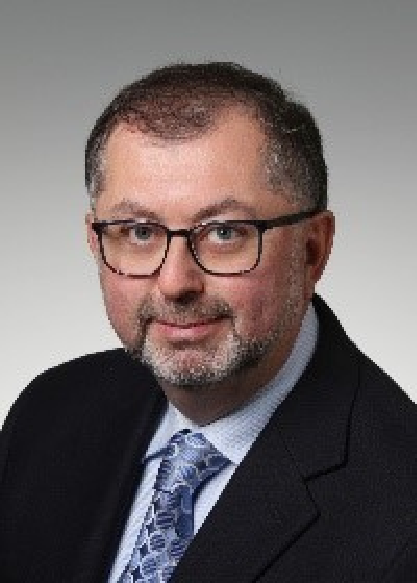}}]{Halim Yanikomeroglu}
(Fellow, IEEE) received the BSc degree in electrical and electronics engineering from the Middle East Technical University, Ankara, Turkey, in 1990, and the MASc degree in electrical engineering (now ECE) and the PhD degree in electrical and computer engineering from the University of Toronto, Canada, in 1992 and 1998, respectively. Since 1998 he has been with the Department of Systems and Computer Engineering at Carleton University, Ottawa, Canada, where he is now a Chancellor's Professor.

Dr. Yanikomeroglu’s research interests cover many aspects of wireless communications and networks, with a special emphasis on non-terrestrial networks (NTN) in the recent years. He has given 110+ invited seminars, keynotes, panel talks, and tutorials in the last five years. He has supervised or hosted over 160 postgraduate researchers in his lab at Carleton. Dr. Yanikomeroglu’s extensive collaborative research with industry resulted in 40 granted patents. Dr. Yanikomeroglu is a Fellow of the IEEE, the Engineering Institute of Canada (EIC), the Canadian Academy of Engineering (CAE), and the Asia-Pacific Artificial Intelligence Association (AAIA). He is a Distinguished Speaker for the IEEE Communications Society and the IEEE Vehicular Technology Society, and an Expert Panelist of the Council of Canadian Academies (CCA|CAC).
Dr. Yanikomeroglu is currently serving as the Chair of the Steering Committee of IEEE’s flagship wireless event, Wireless Communications and Networking Conference (WCNC). He is also a member of the IEEE ComSoc Governance Council, IEEE ComSoc GIMS, IEEE ComSoc Conference Council, and IEEE PIMRC Steering Committee. He served as the General Chair and Technical Program Chair of several IEEE conferences. He has also served in the editorial boards of various IEEE periodicals.

Dr. Yanikomeroglu received several awards for his research, teaching, and service, including the IEEE ComSoc Fred W. Ellersick Prize (2021), IEEE VTS Stuart Meyer Memorial Award (2020), and IEEE ComSoc Wireless Communications TC Recognition Award (2018). He received best paper awards at IEEE Competition on Non-Terrestrial Networks for B5G and 6G in 2022 (grand prize), IEEE ICC 2021, IEEE WISEE 2021 and 2022.
\end{IEEEbiography}

\vfill

\end{document}